\documentclass[manuscript]{aastex}
\usepackage{emulateapj5}

\shorttitle{The distance to CTB109}
\shortauthors{Kothes, Uyan{\i}ker, and Yar}

\begin{document}

\title{The distance to the SNR CTB109 deduced from its environment}
\author{Roland Kothes\altaffilmark{1,2}, B\"ulent Uyan{\i}ker\altaffilmark{1}, 
and Aylin Yar\altaffilmark{1}}
\altaffiltext{1}{National Research Council of Canada,
              Herzberg Institute of Astrophysics,
              Dominion Radio Astrophysical Observatory,
              P.O. Box 248, Penticton, British Columbia, V2A 6K3, Canada}

\altaffiltext{2}{Department of Physics and Astronomy, University of Calgary,
             2500 University Drive N.W., Calgary, AB, Canada}

\email{roland.kothes@nrc.ca, bulent.uyaniker@nrc.ca, aylin.yar@nrc.ca}

\begin{abstract}
We conducted a study of the environment around the supernova remnant
CTB109. 
We found that the SNR is part of a large complex of \ion{H}{2}
regions extending over an area of 400~pc along the Galactic
plane at a distance of about 3~kpc at the closer edge of the
Perseus spiral arm. At this distance CTB109 has a diameter
of about 24~pc. 
We demonstrated that including
spiral shocks in the distance estimation is an ultimate requirement
to determine reliable distances to objects located in the
Perseus arm. The most likely explanation
for the high concentration of \ion{H}{2} regions and SNRs is that
the star formation in this part of the Perseus arm is triggered by
the spiral shock. 
\end{abstract}


\section{Introduction}

The supernova remnant (SNR) CTB109 was discovered by \citet{gregory}
at X-rays with the Einstein Satellite. It is believed that the
peculiar semicircular shape of the SNR results from its interaction
with a giant molecular cloud \citep{tat87} which inhibits its
expansion to the west. It has an unusual double-shell structure in
radio, with an inner half-shell surrounded by a wider outer shell;
both shells have similar radio surface brightness at 1420~MHz (see
Fig. \ref{all4}). CTB109 was classified as a shell-type SNR based on
its radio properties \citep{downes}.  The supernova explosion left a
pulsar behind that is observable in X-rays \citep{fahlman} but no
radio emission has been detected from the pulsar and there is no
evidence for a pulsar wind nebula.

Establishing the distance to a SNR is usually quite difficult, and
CTB109 seems to offer a particular challenge. $\Sigma$~-~D relations
were used by \citet{sofue} and \citet{hughes} to derive distances of
4.1~kpc and 5.6~kpc respectively.  The newest approach to $\Sigma$~-~D
is that of \citet{case}, who use only shell-type SNRs; this gives a
distance of 4.7~kpc. However, the $\Sigma$-D technique, while the
oldest, is probably the most disputed method of distance determination
for SNRs. There are many statistical studies which show that there
is no evidence for a working $\Sigma$-D relation
\citep{green,berkhuijsen86,berkhuijsen87}.

The spectroscopic distance of \ion{H}{2} regions close to CTB109
offers an alternative approach, and \citet{tatematsu} quote values of
3.6~-~5.4~kpc . However, these distances are based on very old
measurements and only two nearby \ion{H}{2} regions were used in this
publication.  Despite the results mentioned above, all previous
authors have used a distance of 4~kpc for CTB109.

The primary motivation for the present study is to determine a more
reliable distance to this SNR. We use spectroscopic distances and
radial velocities of 16 nearby \ion{H}{2} regions and compare these
values with those obtained for CTB109. We also include a comparison of
absorption profiles of the radio bright \ion{H}{2} regions with
foreground \ion{H}{1} column densities of CTB109 calculated from X-ray
absorption and discuss the cold environment of the SNR based on
high-resolution \ion{H}{1} and CO measurements from the Canadian
Galactic Plane Survey (CGPS).

We have also obtained 408~MHz and 1420~MHz radio continuum data
including polarization at 1420~MHz. A joint spectral index and
polarization analysis of these data in conjunction with 2.7~GHz,
4.85~GHz, and 10.55~GHz measurements from the Effelsberg 100~m radio
telescope will be published elsewhere \citep{kothes}.

\section{Observations and Data Analysis}

The CGPS is described in detail by \citet{taylor}. To study the
environment of the SNR CTB109 we have used parts of the CGPS which
derive from observations with the Synthesis Telescope of the Dominion
Radio Astrophysical Observatory (DRAO) \citep{landecker} and CO
observations \citep{heyer} from the Five College Radio Astronomy
Observatory (FCRAO).  For the DRAO data angular resolution varies as
cosec(declination) and therefore changes slowly across the final maps.
Parameters for all observations used can be found in Tab. \ref{param}.
To ensure accurate representation to the largest scales for the DRAO
data we have incorporated single-antenna observations after suitable
filtering in the Fourier domain.  Single-antenna \ion{H}{1} data were
obtained from a survey of the CGPS region made with the DRAO 26-m
Telescope \citep{higgs}. The single-antenna data for the 1420~MHz
radio continuum were obtained from the Effelsberg 1420~MHz Survey of
the Galactic Plane \citep{reich21}.

\section{The distance to CTB109}

In light of all the distance problems around CTB109, we have embarked
on a more thorough analysis and have tried to determine a more
accurate distance to this SNR. Abandoning the $\Sigma$-D relation, the
most commonly used method to determine the distance to a SNR is to
deduce the radial velocity of associated neutral hydrogen or molecular
material. A Galactic rotation model is then used to assign to the
radial velocity a distance from the Sun. The molecular material
associated with CTB109 has a radial velocity of about $-$51~km/s
\citep{tatematsu} (see also section 3.2), implying a kinematic
distance of 5~kpc (using a flat rotation model with $v_\odot$ =
220~km/s and $R_\odot$ = 8.5~kpc).  However, the validity of the flat
rotation model has been challenged for the Perseus arm. \citet{roberts}
pointed out that \ion{H}{2} regions show a high discrepancy between
their kinematic and spectroscopic distances. Furthermore, O-type stars
in the Perseus arm region often show interstellar absorption lines,
arising of course in material in front of them, with velocities more
negative than the radial velocity of the star itself. According to the
flat rotation model this material should be behind those
stars. \citet{roberts} successfully modeled these discrepancies with
the two-armed spiral shock (TASS) model.

In Tab. \ref{hiilist} we list parameters of 16 Sharpless \ion{H}{2}
regions between Galactic longitudes $104^\circ$ and $114^\circ$ taken
from \citet[and references therein]{brand}, an extensive study of the
velocity field of the outer Galaxy. In Fig. \ref{hiiplot} the radial
velocity of these \ion{H}{2} regions is plotted as a function of their
spectroscopic distances. There are two groups of \ion{H}{2} regions,
separated by their velocity and distance. The first group comprises
sources with distances well below 2~kpc and radial velocities between
0 and $-$20~km/s. These sources are located in the Orion Spur, and
their spectroscopic and kinematic distances agree quite well.  There
is a large gap which corresponds to the interarm region between the
Orion Spur and the Perseus arm. The \ion{H}{2} regions of the second
group are all located in the Perseus arm, and most are much closer
than predicted by the flat rotation curve.  This result fits well with
Roberts' predictions \citep[Fig. 4]{roberts}. According to Roberts'
model, the radial velocity in the direction of CTB109 would drop to
about $-$50 to $-$55~km/s at the position of the spiral shock at a
distance of about 2.5~kpc.  It then slowly rises until it rejoins the
flat rotation curve at about 3.5~kpc. The only \ion{H}{2} regions with
velocities between $-$40 and $-$60 km/s whose spectroscopic distances
exceed 4~kpc are the Sh147/8/9 complex and Sh156 (and these two
distances have the biggest errors). Independent distance estimates for
these two \ion{H}{2} regions obtained from infrared measurements by
\citet{wouterloot} are about 3.5~kpc for both.  The distance
ambiguity for Perseus arm velocities is also indicated by the presence
of many \ion{H}{1} self absorption features (HISA), found by
\citet{gibson}.  The HISA phenomenon requires warm neutral gas behind
cold absorbing gas.  The emission of the background hydrogen is then
absorbed by colder foreground material at the same radial
velocity. This can only occur when two distances correspond to the
same radial velocity: a flat rotation curve does not permit this in
the outer Galaxy (e.g. Fig. \ref{hiiplot}), but the Roberts TASS model
does.

We have also produced absorption profiles for the radio-bright
\ion{H}{2} regions in our list. From these we were able to calculate
accurate foreground \ion{H}{1} column densities, also listed in
Tab. \ref{hiilist}. These values are comparable to those obtained by
\citet{rho} towards CTB109 from X-ray absorption. They found absorbing
\ion{H}{1} column densities between 8 and $10\cdot 10^{21}$~cm$^{-2}$.
\citet{patel} derived an absorbing \ion{H}{1} column density of N$_H =
9.3\cdot 10^{21}$~cm$^{-2}$ for the central pulsar from Chandra
observations. For the radio-bright southern spot which is absorbed by
foreground material in our data we measure an absorbing \ion{H}{1}
column density of $8.4\pm 0.8 \cdot 10^{21}$~cm$^{-2}$.

The absorption profiles obtained for the radio bright southern spot of
CTB109, the two closest \ion{H}{2} regions Sh149 and Sh152, and a
nearby extragalactic source are plotted in Fig. \ref{abs}. All
profiles are very similar for the local gas. The profiles of the three
Galactic objects are also very similar for the Perseus arm gas,
indicating that they are located at comparable distances. However, the
extragalactic source has an additional component around $-$45~km/s
which is completely missing in all of the Galactic sources (and so
must arise beyond the Galactic sources).  This is another
indication for the existence of the TASS, because the
radial velocity within the Perseus arm would be increasing with
distance in the TASS model while it would be decreasing if the flat
rotation curve applied.  This evidence also favors the closer distance
for Sh 149, since material with a radial velocity of around -45~km/s,
which is absorbed by the extragalactic source but not by Sh 149,
should be located at around 4~kpc according to \citet{roberts}.  We
should note at this point that the deduction of the systemic velocity
of the Galactic sources from their absorption spectra is not possible
due to the peculiar behavior of the rotation curve in the Perseus
arm.

CTB109 contains a pulsar, discovered by \citet{fahlman}, and evidently
is the product of the explosion of a massive star. Hence it seems
quite reasonable to associate it with the other massive stars in the
vicinity, which are exciting the nearby \ion{H}{2} regions.  In light
of all this new information we propose that CTB109 and the nearby
\ion{H}{2} regions are part of the same complex, which is located in
the shock zone of the Perseus arm as described by \citet{roberts}.
This would imply a distance of $3.0\pm 0.5$~kpc.  The mean distance of
all \ion{H}{2} regions listed in Tab.~\ref{hiilist}, weighted by
1/$\sigma^2$, is $3.1\pm 0.2$~kpc, which supports the above mentioned
distance estimate. The linear size of the SNR at 3.0~kpc would be
24~pc.  The existence of this group of \ion{H}{2} regions at about the
same distance suggests that star formation in the Perseus arm in this
vicinity was triggered by the spiral shock.

\section{The cold environment of CTB109}

The 1420~MHz radio continuum image of CTB109 taken from the CGPS is
shown in Fig. \ref{all4}. The radio emission shows two shells of
radius $8\arcmin$ and $18\arcmin$.  Both shells are incomplete in the
northwest where the radio emission abruptly stops.  The binary pulsar
1E 2259+586 lies near the geometric center of the larger shell.  There
is another bright emission structure in the south where both shells
overlap.  The remnant is brightest to the northeast and southeast with an
intervening depression. A striking feature of the radio emission is
the presence of several thin filaments between the shells and even
some on top of them, indicating complex dynamical structures within
the remnant and around it.

\subsection{The molecular material}

The molecular environment of CTB109 was studied intensively by
\citet{tat87,tatematsu} on the basis of CO observations
with the Nagoya 4-m radio telescope and the 45-m
telescope at the Nobeyama Radio Observatory, respectively. They described the massive
molecular cloud to the west of CTB109 and found a ridge of molecular
material extending inside the SNR, showing an anti-correlation with
the X-ray emission. They found no evidence of shocked CO inside or
around the remnant.  In Fig. \ref{co} we plot channel maps of our CO
data towards CTB109 around the systemic velocity of the SNR.  We
should note at this point that, in contrast to the customary Perseus
arm velocity profile, in this vicinity higher negative velocity
indicates a shorter distance (see also section 3).

The massive molecular cloud to the west of the SNR is very prominent
in the channel maps. This cloud blocks the shock wave from further
expansion in this direction. According to \citet{tat87} this cloud is
responsible for the peculiar shape of the SNR. In the northern part
there is an extension from the cloud into the SNR separating both
radio shells. This CO ridge appears at about $-$46~km/s to $-$47~km/s
and vanishes at $-$52~km/s while the main body of the cloud appears at
a slightly more negative velocity, moving in gradually from the west
until it disappears in the same direction at $-$54~km/s to $-$55~km/s. This
indicates that the main body of the cloud is located closer to us than
the ridge, or that the CO ridge is moving away from us relative to the
big cloud.  It also shows that the core of the massive cloud is to the
west and the SNR shock wave has been reflected by its eastern edge.  The
dark cloud at the tip of the ridge, which is more pronounced in the
integrated map in Fig. \ref{all4}, causes strong absorption in the
X-ray data \citep{tat87,tatematsu,rho} indicating that it lies in front of the
SNR or is associated with it. The only possible radial velocity for 
CTB109 would be between $-$50~km/s and $-$52~km/s where both structures, 
the ridge and the massive cloud, are present. This also
explains the weak outer radio shell beyond the CO ridge. The shock
wave traveling in that direction has mostly been blocked by this ridge.

\subsection{Neutral hydrogen}

Associated neutral hydrogen features are less pronounced than those
seen in CO, but the general structure seen in the CO data is also
present. The massive cloud to the west extends further to the south
and is generally closer to the remnant. The ridge in the north is also
visible, although it is seen slightly further south, closer to the
inner SNR shell, and can be followed further around the inner shell.
It appears to be smoother, which makes it difficult to detect in the
channel maps (Fig. \ref{hi}), but in the integrated map of
Fig. \ref{all4} it is quite obvious. The SNR seems to be located
at a density gradient in the \ion{H}{1} distribution most prominent
in the three channel maps in the second row of Fig.~\ref{hi}. This gradient goes
from high densities in the west smoothly down to a hole in the \ion{H}{1}
distribution east of CTB109. Fig. \ref{all4} also reveals a
diffuse shell of HI surrounding almost the whole outer radio shell of
the SNR. Note that the deep depression in the area between
the bright spot in the south and the northeastern part of the outer
shell (in the \ion{H}{1} channel maps between -47 and -53 km/s)
cannot be caused by absorption since the continuum emission of
the outer shell in this area is not strong enough. Therefore this
relative absence of \ion{H}{1} emission is due to the presence of the
smooth outer shell. 
Absorption of the bright southern spot is also detectable in the
channel maps.

\section{Discussion}

\subsection{CTB109 as part of a large complex}

The distribution of \ion{H}{2} regions and SNRs in the vicinity of
CTB109 is displayed in Fig.~\ref{longdist} in relation to the
Perseus arm locations from \citet{roberts} and \citet{taylorcordes}.
Apparently the kinematic distance for CTB109
derived from the flat rotation curve would place the SNR in the interarm 
region behind the Perseus arm. Even the 4~kpc distance assumed by 
various earlier authors, would be outside the spiral arm. Since the
SNR is associated with dense molecular and atomic material an interarm
location would be very unlikely. 

In the area shown in Fig.~\ref{longdist}
there are two concentrations of sources within the Perseus arm. 
The first is around CTB109 and the supernova remnant
Cas A and the second is around the Tycho SNR. There is a third
concentration at higher longitudes around the W3/4/5 complex and
the supernova remnant HB3 (not shown here). Obviously, while traveling 
through the Perseus arm the spiral shock triggered star formation in 
high-density regions, where we now find these concentrations of young 
stellar objects, indicated by the compact \ion{H}{2} regions and 
supernova remnants which are the result of the explosion of massive 
young stars as well. 

\subsection{Neutral material in the vicinity of CTB109}

CTB109 is located at a density gradient which goes from very high densities
on the molecular cloud in the west to a gap in the emission in
the east of the SNR. Apparently the progenitor
star was formed at the edge of a dense molecular cloud while the star
in Sh 152 was formed in its centre. To the west, the shock wave 
expanded into
the dense cloud which decelerated it very quickly. To the east, however,
the shock wave is expanding into a moderately dense medium. The structure
of the surrounding \ion{H}{1} indicates that the SNR is not expanding
inside a stellar wind bubble, because we would expect such a structure
to have a more pronounced outer boundary. It is more likely that the missing
emission from the interior is the result of taking away the neutral 
hydrogen by ionizing it with the expanding shock wave. This would imply
a progenitor star of type B2/3 since more massive stars would have a strong
stellar wind.

Between the outer and inner shell in the
north we find the ridge of cold material separating both structures. The
fact that the \ion{H}{1} is smoother and mostly concentrated outside
the thin, dense molecular ridge indicates an interaction with the SNR in
which the surface of the molecular structure was dissociated or even
evaporated by the expanding shock wave. The lack of X-ray emission
coinciding with the ridge indicates that the ridge is located on the 
near side of the remnant and is absorbing the 
X-ray emission which originates behind it.

The outer radio shell shows two very prominent features. These are the 
bright knot to the south and the bright part of the northeastern shell.
Both features are located within bright parts of the surrounding \ion{H}{1}
(see Fig.~\ref{all4}) indicating that these parts of the remnant 
are expanding 
into a higher density medium than the part between them where the emission 
of the radio shell is 
weakest and the surrounding \ion{H}{1} seems to have almost a gap. This
suggests a close relation between the radio brightness of the SNR's expanding
shell and the density of the medium it is expanding into, which is of course 
expected.

\section{Summary}

We have presented new radio continuum data for the SNR CTB109 together
with \ion{H}{1} and CO observations of the surrounding medium. Our
data show that the radio continuum emission from the remnant is
closely related to the surrounding cold material. Analysis of the
\ion{H}{1} and CO dynamics and comparison of the results with
parameters of nearby \ion{H}{2} regions has led us to the conclusion
that the SNR is located at a distance of $3.0\pm 0.5$~kpc, as opposed
to larger distances previously published in the literature. It also
implies that the SNR, together with several \ion{H}{2} regions, is
part of a large complex created by the spiral shock present in the
Perseus arm.  We have shown that inclusion of the spiral shock is
necessary to produce reliable kinematic distances in this part of the
Galaxy, and we have demonstrated that, in seeking the distance to a
supernova remnant, it is important to take a broader view of the
environment. The wide-field, high-resolution data of the CGPS,
revealing multiple components of the ISM, are ideal for a study of
this type.

\acknowledgments
We like to thank Tom Landecker for careful reading of the
manuscript. We also like to thank Christopher Brunt for many useful
discussions about the spiral shock in the Perseus arm.
The Dominion Radio Astrophysical Observatory is a National Facility
operated by the National Research Council.  The Canadian Galactic
Plane Survey is a Canadian project with international partners, and is
supported by the Natural Sciences and Engineering Research Council
(NSERC). 

%\appendix

\clearpage

\begin{deluxetable}{ll}
\tablewidth{0pc}
\tablecolumns{2}
\tablecaption{CGPS Data Parameters at the position of CTB109 \label{param}}
\tablehead{
\colhead{Parameter} & \colhead{Value}
}
\startdata
Pixel Size                   &  $18\arcsec$ \\
Number of Spectral Channels  &  256 \\
Channel Separation           & 0.8243 km s$^{-1}$ \\
LSR Velocity Range           & $+45$ km/s to $-170$ km/s  \\
\ion{H}{1} Spectral Resolution          & 1.319 km/s \\
CO Spectral Resolution & 1.0 km/s\\
\ion{H}{1} Beam Size\tablenotemark{a}   & $68\arcsec \times 59\arcsec$ at $-63\degr$ \\
CO Beam Size  & $45'' \times 45''$ \\
\ion{H}{1} RMS Noise per Channel        &  2.5 K   \\
CO RMS Noise per Channel        &  0.2 K   \\
\enddata
\tablenotetext{a}{Position angle of the beam is with respect to the
Galactic pole.}
\end{deluxetable}

\begin{table}
  \caption{Parameters of 16 nearby \ion{H}{2} regions taken from \citet{brand}.
  The errors in the \ion{H}{1} column density values are between 10 and 20~\%.}
  \label{hiilist}
 \begin{center}
  \begin{tabular}{lcr@{$\pm$}lr@{$\pm$}lc} \hline \hline
  & & \multicolumn{2}{c}{ } & \multicolumn{2}{c}{ } & \\
  Name & Gal.-Coor. & \multicolumn{2}{c}{Distance [kpc]} & 
  \multicolumn{2}{c}{v$_{LSR}$ [km/s]} & N$_{HI}$ [$10^{21}$~cm$^{-2}$] \\
  & & \multicolumn{2}{c}{ } & \multicolumn{2}{c}{ } & \\ \hline
  & & \multicolumn{2}{c}{ } & \multicolumn{2}{c}{ } & \\
  Sh135 & 104.59+1.37 & 1.4 & 0.4 & -20.7 & 0.5 & - \\
  Sh137 & 105.15+7.12 & 0.6 & 0.2 & -10.3 & 1.4 & - \\
  Sh139 & 105.77$-$0.15 & 3.3 & 1.1 & -46.5 & 0.5 & - \\
  Sh140 & 106.81+5.31 & 0.9 & 0.1 & -8.5 & 1.0 & - \\
  Sh142 & 107.28$-$0.90 & 3.4 & 0.3 & -41.0 & 0.5 & - \\
  Sh149 & 108.34$-$1.12 & 5.4 & 1.7 & -53.1 & 1.3 & 8.4 \\
  Sh152 & 108.75$-$0.93 & 3.6 & 1.1 & -50.4 & 0.5 & 8.2 \\
  Sh154 & 109.17+1.47 & 1.4 & 0.4 & -11.5 & 0.9 & - \\
  Sh155 & 110.22+2.55 & 0.7 & 0.1 & -10.0 & 1.5 & - \\
  Sh156 & 110.11+0.05 & 6.4 & 2.0 & -51.0 & 2.0 & 8.7 \\
  Sh157 & 111.28$-$0.66 & 2.5 & 0.4 & -43.0 & 2.0 & -\\
  Sh158 & 111.54+0.78 & 2.8 & 0.9 & -56.1 & 1.1 & - \\
  Sh159 & 111.61+0.37 & 3.1 & 1.2 & -56.0 & 1.0 & 11.0 \\
  Sh161B & 111.89+0.88 & 2.8 & 0.9 & -51.9 & 0.7 & - \\
  Sh162 & 112.19+0.22 & 3.5 & 1.1 & -44.7 & 0.5 & 7.7 \\
  Sh163 & 113.52$-$0.57 & 2.3 & 0.7 & -44.9 & 3.8 & - \\ 
  & & \multicolumn{2}{c}{ } & \multicolumn{2}{c}{ } & \\ \hline
  \end{tabular}
\end{center}
\end{table}

\clearpage

\begin{figure}
   \caption{upper left: Radio continuum image of the SNR CTB109 taken from 
   the CGPS at 1420~MHz. The shading goes from 5~K (white) to 18~K (black). 
   The position of the X-ray pulsar 1E2259+586 is marked 
   by a white asterisk. Upper right: CO around CTB109 integrated between -45~km/s and 
   -55~km/s. Contours are from 0.2~K (white) to 4.2~K (black) in steps of 0.5~K. 
   Lower left: \ion{H}{1} around CTB109 integrated between -45~km/s 
   and -55~km/s. Contours are from 70~K (black) to 94~K (white) in steps of 3~K.}
   \label{all4}
\end{figure}

\begin{figure}
   \caption{Radial velocities of \ion{H}{2} regions in the vicinity
   of CTB109 as a function
   of their spectroscopic distances. The solid line represents the
   expected behaviour according to the flat rotation model with
   $v_\odot$ = 220~km/s and $R_\odot$ = 8.5~kpc. The radial velocity
   of CTB109 is indicated by the dashed line.}
   \label{hiiplot}
\end{figure}

\begin{figure}
   \caption{Absorption and emission profiles of the southern bright spot of 
   CTB109, Sh149, Sh152, and the extragalactic radio source KR~68a.}
   \label{abs}
\end{figure}

\begin{figure}
   \caption{CO channel maps towards the SNR CTB109 taken from
   the CGPS. White contours indicate the radio continuum emission at 1420~MHz.}
   \label{co}
\end{figure}

\begin{figure}
   \caption{\ion{H}{1} channel maps towards the SNR CTB109 taken from
   the CGPS. White contours indicate the radio continuum emission at 1420~MHz.}
   \label{hi}
\end{figure}

\begin{figure}
   \caption{\ion{H}{2} regions (open circles) and SNRs (open stars) in the
   vicinity of CTB109. The Perseus arm as defined by \citet{roberts}
  is indicated by the dashed lines. The location of the maximum electron 
   density in the Perseus arm as determined by \citet{taylorcordes} is marked
   by the dotted line. The names of the individual SNRs are noted.
   For Sh156 and the Sh147/8/9 complex the distances from
   \citet{wouterloot} are used and for all other \ion{H}{2} regions the
   spectroscopic distances listed in \citet{brand}. The distances for the SNRs Cas~A and Tycho were
   taken from \citet{reed} and \citet{chevalier}, respectively.} 
   \label{longdist}
\end{figure}

\end{document}